\documentclass[twocolumn,prl,floatfix,citeautoscript,nofootinbib,superscriptaddress]{revtex4}
\usepackage{amsbsy}
\usepackage{latexsym,epsfig,graphicx}
\usepackage{dcolumn}
\usepackage{graphicx}
\usepackage{subfigure}
\usepackage{comment}
\usepackage{color}
\usepackage{bm}
\usepackage{mathrsfs}
\usepackage{amsfonts}
\usepackage{amsmath}
\usepackage{color}
\usepackage{amssymb}
\usepackage{xspace}
\usepackage{epstopdf}
\usepackage{tabularx}
\usepackage{longtable}
\usepackage[colorlinks=true, letterpaper=true, pdfstartview=FitV, linkcolor=red, citecolor=blue, urlcolor=blue]{hyperref}
\usepackage[normalem]{ulem}
\begin{document}
\title{Dicke-Ising quantum battery of an ion chain driven by a mechanical oscillator }
\author{Jun Wen} 
\affiliation{School of Mathematics, Physics and Statistics, Sichuan Minzu college, Ganzi 626001, China}
\author{Zheng Wen} 
\affiliation{Office of Physics and Chemistry, Army Academy of Border and Coastal Defence, Xi'an 710108, China}
\author{Ping Peng} 
\affiliation{Department of Physics and Institute of Theoretical Physics, Shaanxi University of Science and Technology, Xi'an 710021, China}
\author{Guan-Qiang Li} 
\thanks{Corresponding author email: liguanqiang@sust.edu.cn}
\affiliation{Department of Physics and Institute of Theoretical Physics, Shaanxi University of Science and Technology, Xi'an 710021, China}

\begin{abstract}
A scheme for implementing quantum batteries in a realizable and controllable platform based on a trapped ion chain
driven by a mechanical oscillator is proposed. The effects of the hopping interaction between the two-level ions and the coupling interaction between the ions and the external mechanical oscillator on the charging process of the battery are investigated. The importance of the counter-rotating wave terms in the system's Hamiltonian, which are often ignored, is analyzed, and it is found that the charging energy and the ergotropy of the battery are dramatically affected by the counter-rotating wave terms. The quantum phase transition of the two-level system is restrained by the counter-rotating wave terms due to the destruction of the quantum coherence. Lastly, the power-law dependence of the charging process on the distance between the ions is discussed. Our theoretical analysis provides a solid foundation for the development of a practical quantum battery.
\end{abstract}
\maketitle

\section{I.~Introduction}
The battery is an important device that can store, deliver and convert energy. Such device has produced a huge impact on modern civilization and attracted great enthusiasm for researches since its invention. With the development of the science and technology, more and more researchers have been trying to apply the quantum theories and techniques to the battery for improving its performance. In the process of the development, the concept of the quantum battery was put forward~\cite{F.Campaioli2023}. In 2013, the fundamental theory of the quantum battery's charging was first established and the relationship between the passive states, the ergotropy and the Gibbs canonical density matrix is investigated by R. Alicki and M. Fannes~\cite{RAMF}  based on the work given in Ref.~\cite{Allahverdyan2004}. In the same year, a method at the price of time to extract all possible work without creating any entanglement was provided, and the relation between the entanglement and the ergotropy was discussed \cite{KVHM}. In 2015, the protocol for charging the single qubit was demonstrated and then extended to the system containing $N$ qubits~\cite{FCBS}. F. Campaioli et.al pointed out that the charging power would be enhanced if $N$ batteries are charged collectively \cite{FCFA}. In 2018, it has been found that the anisotropic interactions for the spin chain can provide a boost to the charging power, but the analysis was done under the condition of the weak interaction~\cite{TPLJ}. G. M. Andolina et.al studied the reason of the fast charging of quantum batteries, including the effects of many-body interaction and quantum-mechanical nature of the model itself, the result indicated that it depends on the specific models and parameters~\cite{GMAM}. In 2020, it is pointed out that the quantum coherence in the battery or the battery-charger entanglement is a necessary resource for generating nonzero extractable work during the charging process~\cite{W.L.Yang2022}. Optimal control methods are used for preparing quantum states and recently they are exploited for fast charging quantum batteries~\cite{F.Mazzoncini2023, RRRodriguez2024}.

In addition, the realistic quantum system is affected by environment inevitably, and an increasing number of research works are devoting to the study of quantum batteries under the framework of open systems~\cite{KMolmer2019,J.Liu2019,MCarrega2020}. The charging problem of quantum batteries in non-Markovian environments shows that the memory effect of the environments can allow the energy in the battery to be present for a longer period of time~\cite{FHKF}. The effect of environmental decoherence can be effectively suppressed in the presence of two Floquet states in the quasienergy spectrum and the aging of the battery can be extended~\cite{SYBJ}. Using the thermal reservoir as an energy source, it is found that the multiple coupled spin bits can convert thermal energy into useful work for charging of the quantum battery, which is the opposite of what happens with a single bit~\cite{FZFQ}. Furthermore, the significance of the presence of a catalyst system between the battery and the charger for charging has also been investigated~\cite{RRRB}. The explorations above and other related works (such as~\cite{D.Rosa2022-1,D.Rosa2020-2,Downing2023,C.Shang2024}) have laid foundation for the further investigation of the practical quantum batteries.

Inspired by the recent progress, a question arises: is there a realistic platform, which is
easy to be prepared and manipulated, to develop the quantum batteries?
At present, several schemes of quantum batteries based on superconductors~\cite{CKHu2022}, quantum dots~\cite{Wenniger2022}, organic microcavities~\cite{JQQuach2022} and nuclear spin~\cite{JJoshi2022} have been proposed, but they are still far from practical applications. The ion trap, which has been thought as an excellent platform, has demonstrated many important applications in the quantum fields, including quantum phase transition \cite{XLDD,DPJI}, quantum simulation~\cite{RBCF,Duan2024} and quantum computation~\cite{KKim2019,C.Noel2022}. In particular, the study done by Duan and coworkers has realized a quartic trapping potential and allowed one to trap and control 512 ions experimentally~\cite{Duan2024}.  How to realize the quantum batteries in the platform of ion trap? The technique of the quantum optics allows us to regulate and control the trapped ions accurately~\cite{XLDD}, which is the basis of realizing the quantum batteries of trapped ion chains.

In the present paper, a practical proposal to implement the quantum battery based on a trapped ion chain is shown. The ions trapped in a Paul trap are coupled with a mechanical harmonic oscillator, as shown in Fig.~1. The pump field is used to drive the oscillator. The coupling between the mechanical mode and the spin due to the gradient magnetic field is tunable precisely \cite{PBLY}. The magnetic field is produced by a magnetic film deposited on a cantilever with nanoscale thickness. The system can transfer the energy of phonons to the quantum battery by the interactions of the spin and mechanical degree of freedom. The $N$ two-level ions are assumed to couple the external mechanical oscillator with same strength. We investigate the charging process of the trapped ion quantum battery affected by the hopping interaction between the ions and the coupling interaction between the ions and the mechanical oscillator, discuss the effect of the counter-rotating wave terms of the system's Hamiltonian on the process, and obtain the maxima of the charging energy and ergotropy under certain situations. It is hoped that our research can provide a reference for the development of the realistic quantum batteries.

The rest parts of the paper are structured as follows. In Sec.~II, we introduce the
Hamiltonian of our charging system, define some important physical quantities to characterize the
charging process of the system. Sec.~III gives the numerical results and explains
the physics behind the phenomena. A brief conclusion is given in the last section.
\begin{figure}[tbph]
\centering\includegraphics[width=9 cm]{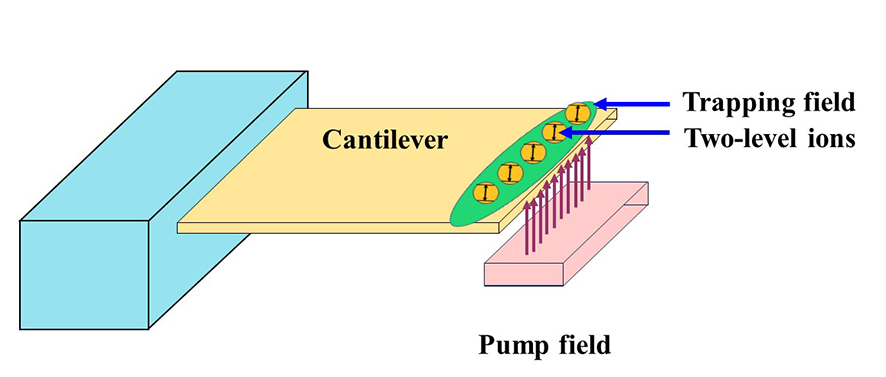}
\caption{(Color online) Schematic diagram of a quantum battery based on a trapped ion chain coupled with a mechanical harmonic oscillator (the cantilever that is fixed at one end and suspended at the other). The pump field is used to drive the oscillator  and the coupling parameter $\lambda$ between the ion chain and the oscillator is tunable precisely. The green region represents the trapping field for the ion train. }
\label{figure1}
\end{figure}

\section{II.~Model and characterizing method}
Fig.~\ref{figure1} is the sketch of our charging system. An ion chain with $N$ two-level ions, in which the level splitting of each ion is $\omega _{a}$, is coupled by a mechanical harmonic oscillator with the angular frequency $\omega _{c}$. The coupling strength between the chain and the oscillator is presented by a controllable parameter $\lambda$. The hopping interaction between the different ions is proportional to $J/\left\vert z_{m}-z_{n}\right\vert^{p}$, in which $z_{m}$ and $z_{n}$ are the scaled equilibrium positions for the $m$th and $n$th ions, respectively. The value of the tunable exponent $p$ is chosen at the range $[0, 3]$~\cite{XLDD,PJBP}. The Hamiltonian of the system is
\begin{eqnarray}
\hat{H} &=&\hat{H}_{c}+\hat{H}_{a}+\hat{H}_{ac},
\notag \\
\hat{H}_{a} &=&\omega _{a}\sum_{n=1}^{N}\hat{\sigma} _{n}^{+}\hat{\sigma}
_{n}^{-}+J\sum_{n=1}^{N}\sum_{m>n}^{N}\frac{\hat{\sigma} _{n}^{x}\hat{\sigma} _{m}^{x}}{%
\left\vert z_{m}-z_{n}\right\vert ^{p}},  \label{1}
\notag \\
\hat{H}_{c} &=&\omega _{c}\hat{c}^{\dag}\hat{c},~~\hat{H}_{ac}=\lambda\sum_{n=1}^{N}(\hat{c}+\hat{c}^{\dag})\hat{\sigma} _{n}^{x}.
\end{eqnarray}
Here, $\hat{\sigma} _{n}^{+}\hat{\sigma} _{n}^{-}=(I_{n}+\hat{\sigma} _{n}^{z})/2$ with $\hat{\sigma} _{n}^{\pm }=(\hat{\sigma} _{n}^{x}\pm i\hat{\sigma}_{n}^{y})/2$. $\hat{\sigma}_{n}^{\alpha }$ and $I_{n}$
are the Pauli and identity matrices for the $n$th ion, where $n=1, 2, ..., N$ and $\alpha =x, y, z$. $\hat{H}_{c}$ is the Hamiltonian of the mechanical harmonic oscillator, $\hat{c}$ ($\hat{c}^{\dag}$) is the corresponding annihilation (creation) operator of phonon. $\hat{H}_{a}$ includes their own energy of the
two-level ions and the hopping energy between the different ions. $\hat{H}_{ac}$ represents the coupling interaction between the trapped ion chain and the oscillator. The Hamiltonian $\hat{H}$ in Eq.~(\ref{1}) can be regarded as the Dicke-Ising model. We have set $\hbar \equiv1$ in the above Hamiltonian. For convenience but without losing the physics, the parameter $\omega _{c}=\omega_{a}=1.0$ is chosen.

We consider the charging process of the system which
consists of the mechanical harmonic oscillator and five $(N=5)$ two-level ions under
the view of the closed system. Our aim is to construct a charging scheme of the quantum battery which can collect low-frequency vibration energy. The influence of the hopping interaction between the ions and the coupling interaction between the ions and the oscillator on the charging process is investigated, with a particular focus on the significance of the counter-rotating wave terms associated with these two interactions~\cite{DBRA}. The initial state of the two-level ions is the ground state $\left\vert g\right\rangle _{a}$ obtained by numerically diagonalizing $\hat{H}_{a}$. The mechanical oscillator is assumed to be prepared in the superposition of Fock states $\left\vert \Phi \right\rangle _{c}=\sqrt{%
0.6}\left\vert 10\right\rangle +\sqrt{0.4}\left\vert 15\right\rangle$. There have been a lot of theoretical studies that have suggested synthesizing superpositions of Fock states in different systems~\cite{VBergholm2019, PDNation2013, QZYY, MKounalakis2019, H.Tan2014, H.Xie2019, JJing2024}. Experimentally, the multi-phonon Fock states and their superposition states have only been obtained inside a high-overtone bulk acoustic wave resonator as far as we know~\cite{Y.Chu2018}. In the following context, we assume that the initial wavefunction of such coupling system is $\left\vert \Psi(0)\right\rangle=\left\vert \Phi\right\rangle _{c}\otimes$ $\left\vert g \right\rangle _{a}$, and the scaled equilibrium positions $ z_{n}$ of the five trapped ions are $\left[ -1.7429,-0.8221,0,0.8221,1.7429\right]$ in the Paul trap~\cite{DFVJ}. The dynamical evolution for the system is based on
\begin{equation}
\left\vert \Psi (t)\right\rangle =e^{-i\hat{H}t}\left\vert \Psi (0)\right\rangle, \label{2}
\end{equation}
and $\hat{\varrho}(t)=\left\vert \Psi (t)\right\rangle \left\langle \Psi
(t)\right\vert$. The reduced density operator is $\hat{\varrho}_{a}(t)=Tr_{c}[\hat{\varrho }(t)]$
and the subscript $c$ indicates a partial trace. The total energy of the ion chain can be calculated by $E(t)=Tr[\hat{H}_{a}\hat{\varrho}_{a}(t)]$. The energy obtained
from the mechanical oscillator is $E_{c}(t)=E(t)-E(0)$, which represents the actual charging energy of the battery. Another key quantity is ergotropy defined as $E_{e}(t)=E(t)-\sum_{k}r_{k}(t)e_{k}$ \cite{Allahverdyan2004},
and the eigenvalues $r_{k}(t)$ and $e_{k}$ are determined by
\begin{equation}
\hat{\varrho}_{a}(t)=\underset{k}{\sum }r_{k}(t)\left\vert r_{k}(t)\right\rangle
\left\langle r_{k}(t)\right\vert,~\hat{H}_{a}=\underset{k}{\sum }e_{k}\left\vert
e_{k}\right\rangle \left\langle e_{k}\right\vert.  \label{3}
\end{equation}
Here, $r_{k}(t)$ are arranged in descending order but $e_{k}$ in ascending order. The ergotropy is used to describe the maximal extractable work from the ion chain. The dynamics of the model (\ref{1}) can not be obtained analytically. The Hilbert space of the quantum harmonic oscillator must be cut off for numerical calculations. The dimension of the space is set to $101$ after the space has been truncated, and the eigenstate for the highest energy level is $\left\vert100\right\rangle $. The dimension number of the Hilbert space exceeds four times of the phonon's number for the harmonic oscillator~\cite{DFMC}.

Several other quantities are introduced to characterize the charging process of the system. $\sigma
_{n}$ represents the expectation value of $\hat{\sigma} _{n}^{+}\hat{\sigma} _{n}^{-}$, which is
defined as $\left\langle \hat{\sigma} _{n}^{+}\hat{\sigma} _{n}^{-}\right\rangle $, and $\sigma _{m,n}=\sigma _{m}-\sigma _{n}.$ In order to show the quantum phase transition, we need to introduce $M_{z}=\left\langle
g\right\vert \hat{S}_{z}\left\vert g\right\rangle _{a}/N$ and $O_{z}=\left\langle
g\right\vert \hat{S}_{z}^{2}\left\vert g\right\rangle _{a}/N^{2}$, where $%
S_{z}=\sum_{n=1}^{N}\sigma _{n}^{z}$. The entropy is significant in both classical and quantum physics, and the von Neumann entropy is used to quantify entanglement of the
pure states~\cite{CHBD}. According to the definition, the von Neumann entropy is expressed as
$S(t)=-\sum_{k}r_{k}(t)\log _{2}r_{k}(t)$. It generally requires the dimensions of
the two subsystems are same, but our model does not meet this condition.
The calculated values of the von Neumann entropy for the two subsystems are
equal. So we can use the von Neumann entropy to describe the disorder of the subsystem and the degree of entanglement for the whole system.

\section{III.~Results and their physical interpretations}
\begin{figure}[tbph]
\centering\includegraphics[width=9 cm]{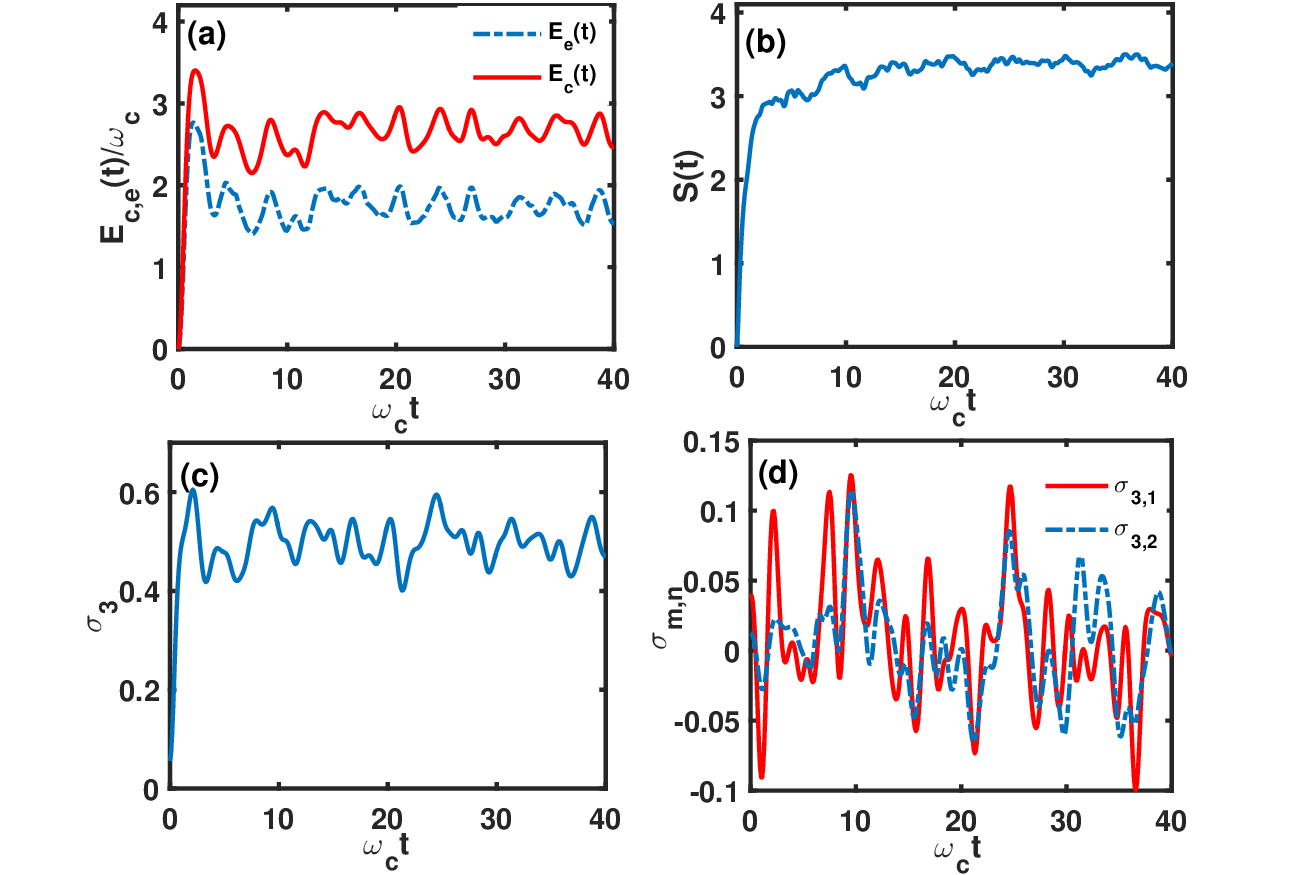}
\caption{(Color online) The charging process of the quantum battery in our model. (a) The evolution of the charging energy $E_{c}(t)$ and the ergotropy $E_{e}(t)$;  (b) The evolution of the entropy $S(t)$;  (c) The dynamics of the expectation value $\left\langle \hat{\sigma} _{3}^{+}\hat{\sigma} _{3}^{-}\right\rangle $ of the third ion; (d) The difference of the expectation value between the different ions. The parameters have been set as $\lambda=0.25$, $J=0.2$ and $p=3$. }
\label{figure2}
\end{figure}

To guarantee that the quantum battery on our platform receives sufficient energy, we have assumed that the
number of the phonons is 12, which is greater than two times of the number of the ions. Actually, the charging process of the quantum batteries is dynamical one, which can be investigated based on Eqs.~(\ref{1}) and (\ref{2}). The quantities for characterizing the process are determined by numerically solving the above equations. Fig.~\ref{figure2} displays the dynamical evolution of the charging energy $E_{c}(t)$, the ergotropy $E_{e}(t)$ and the entropy $S(t)$. In Fig.~\ref{figure2}(a), the evolution of $E_{c}(t)$ and $E_{e}(t)$ gives an irregular oscillation with a small amplitude around an average value after experiencing a relaxation process. The difference between the local maxima and minima (except for $t=0$) for $E_{c,e}(t)$ is relatively small. Such results also hold for the time range $\omega _{c}t\in \lbrack0, 100]$. They are very different from the results in Ref.~\cite{FQDQ} since there exists beating structure in the evolution of $E_{c,e}(t)$ therein. In addition, the difference between the local maxima and minima (except for $t=0$) of $E_{c,e}(t)$ are larger than ours. Both of the counter-rotating wave terms and the non-nearest-neighbour hopping have been considered in our model. For $J=0.2$, the effect of the nearest-neighbor hopping is relatively small, and the effect of the non-nearest-neighbour hopping can be ignored. Therefore, the main reason for the above results is due to the role played by the counter-rotating wave term $\hat{H}_{ac,cw}\equiv$ $\lambda\sum_{n=1}^{N}(\hat{c}^{\dag}\hat{\sigma} _{n}^{+}+\hat{c}\hat{\sigma}_{n}^{-})$ in the Hamiltonian $\hat{H}_{ac}$ for the coupling interaction between the ion chain and the oscillator.

Maybe, some people have a doubt that whether the counter-rotating wave term can play an important role in our system because of $\lambda/\omega _{c}=0.25$ in Fig.~2? In fact, the counter-rotating wave term must be taken into account for $\lambda/\omega _{c}\geq 0.1$ since the system has entered the so-called ultrastrong coupling regime~\cite{XTZH}. Previous studies have demonstrated that the counter-rotating wave term can induce the collapse and revival of the quantum dynamics \cite{JCGR,JFHC}. In 2008, H. Zheng et.al highlighted that the Zeno time is longer by $2$ orders of magnitude than that calculated without considering the counter-rotating wave term \cite{HZSY}. Here, the counter-rotating wave term destroys the quantum coherence and leads to the disappear of the beating structure, and finally restrains the energy exchange between the two-level ions and the mechanical oscillator. In Fig.~2(b), the entropy $S(t)$ rapidly increases from the initial time and then approximately approaches to a stable value. This means the entanglement between the ion chain and the external driving mechanical oscillator is stabilized with the increase of charging time. As an important quantum resource, the change of the entanglement does not show a positive correlation with $E_{c}(t)$ and $E_{e}(t)$ during the charging process. Similar result was also revealed by investigating the (dis-)entanglement via open quantum systems and the Sachdev-Ye-Kitaev battery chargers in Ref.~\cite{U.R.Fischer2024}, showing that the charging power is not an entanglement monotone by computing the negativity measure of entanglement. The dynamics of $\sigma_{3}$, which corresponds to the average energy of the most intermediate ion in the chain, is demonstrated in Fig.~2(c). $\sigma _{m,n}$ indicates that there is the energy difference of the $m$th and $n$th ions although they couple the mechanical oscillator with the same strength, as shown in Fig.~2(d). It indicates that the contribution to the charging process of every ion is different, and the third ion is not superior or inferior although it has a special position in the ion chain.
\begin{figure}[tbph]
\centering\includegraphics[width=7cm]{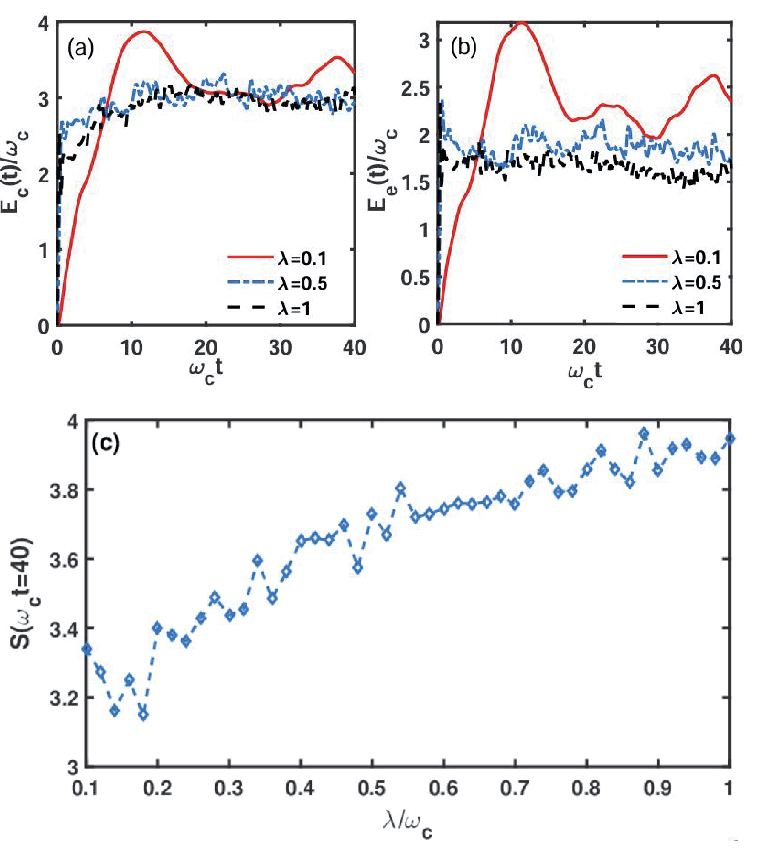}
\caption{(Color online) The comparison of the charging process for the charging energy $E_{c}(t)$ in (a) and the ergotropy $E_{e}(t)$ in (b) under different coupling strength $\lambda$. The change of the final entropy $S(\omega_{c}t=40)$ with $\lambda/\omega_{c}\in[0,1]$ is given in (c). The other parameters are set as $J=0.4$ and $p=3$.}
\label{figure3}
\end{figure}

Fig.~\ref{figure3}(a) and (b) show the evolution of $E_{c}(t)$ and $E_{e}(t)$ under different coupling strength $\lambda$, respectively. Similar to the result for $\lambda=0.25$ and $J=0.2$ given in Fig.~2(a), there exist some irregular oscillations after relaxation at the initial stage. The oscillation amplitude of $E_{c,e}(t)$ is larger than the other two cases corresponding to $\lambda=0.5$ and $\lambda=1.0$ in Fig.~\ref{figure3}(a) since the role played by the counter-rotating wave term $\hat{H}_{ac,cw}$ is relatively small for $\lambda=0.1$. The counter-rotating wave term $\hat{H}_{ac,cw}$ becomes more and more important with increasing $\lambda$. The local maxima of $E_{c}(t)$ and $E_{e}(t)$ have been suppressed heavily for larger $\lambda$. The speed of energy exchange in the process is increased by increasing the coupling strength $\lambda$ and then reducing the oscillation period, which leads to the shorter time to reach the first local maximum. The general tendency for the stable value of $S(t)$ after a long period of time is improved by increasing $\lambda$, as shown in Fig.~3(c) except the value for small $\lambda$. The quantum entanglement between the ion chain and the mechanical oscillator is enhanced correspondingly. But it has no promoting effects for the maximization of $E_{c}(t)$ and $E_{e}(t)$ \cite{DLYang2023}. From the view of the subsystem, the disorder of the two-level system can be described by the entropy $S(t)$. According to the classical thermodynamics, the disorder energy is more difficult to convert into the useful work, so that the ergotropy $E_{e}(t)$ decreases with increasing $\lambda$ shown in Fig.~3(b) can be explained. If the charging time is defined by the time at which $E_{e}(t)$ firstly reaches its local maximum, the charging becomes quicker for larger $\lambda$. So increasing $\lambda$ can improve the charging speed but suppress the ergotropy. A reasonable strength of the coupling interaction between the ions and the oscillator should be taken for the actual battery platforms.
\begin{figure}[tbph]
\centering\includegraphics[width=9cm]{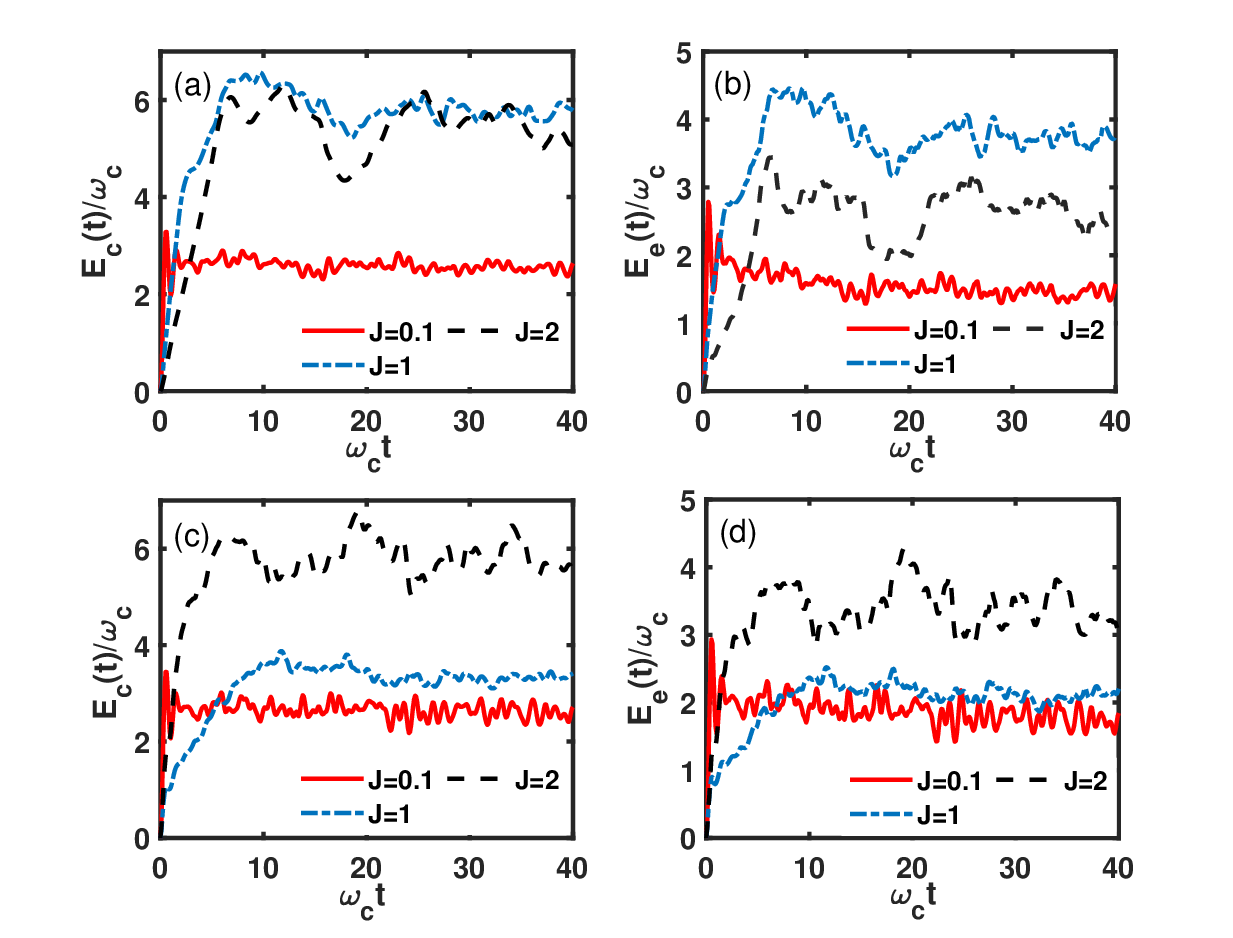}
\caption{(Color online) The comparison of the charging process for the charging energy $E_{c}(t)$ in (a) and the ergotropy $E_{e}(t)$ in (b) under different strength of the hopping interaction $J$. (c) and (d)  present the corresponding results without considering the counter-rotating wave term between the two-level ions. The other parameters are set as $\lambda=0.5$ and $p=3$.  }
\label{figure4}
\end{figure}

Fig.~\ref{figure4}(a) and (b) show the time evolution of $E_{c}(t)$ and $E_{e}(t)$ under different $J$, respectively.
According to the result in Fig.~\ref{figure4}(a) , the final value of $E_{c}(t)$ exceeds about $5\omega _{a}$ (the highest energy of five two-level ions without considering the hopping energy) for larger $J$ after the initial relaxation. As increasing the hopping interaction, the ground energy becomes more lower and the energy spectrum is expanded dramatically [as shown in Fig.~6(a)], which is beneficial for charging and storing more energy in the quantum battery. But for small value of $J=0.1$, the final value of $E_{c}(t)$ is smaller that $3\omega _{a}$. The reason is that only a subset of the ions are excited in this situation, and the contribution from the hopping interaction is insufficient for small $J$. In Fig.~\ref{figure4}(b), the final value of $E_{e}(t)$ for $J=2.0$ is smaller than that for $J=1.0$. That is to say, the final value is not monotonic with increasing $J$. The similar result can also be obtained for $E_{c}(t)$ when $J=1.0$ and $J=2.0$ from Fig.~\ref{figure4}(a). This can be interpreted by the enhancement of the counter-rotating wave term $\hat{H}_{J,cw}\equiv J\sum_{n=1}^{N}%
\sum_{m>n}^{N}(\hat{\sigma} _{n}^{+}\hat{\sigma}
_{m}^{+}+\hat{\sigma} _{n}^{-}\hat{\sigma} _{m}^{-})/\left\vert z_{m}-z_{n}\right\vert
^{3}$ between the ions. Fig.~\ref{figure4}(c) and (d) show the results under the same conditions as in Fig.~\ref{figure4}(a) and (b), but without considering the role of the counter-rotating wave term $\hat{H}_{J,cw}$.
The final values of $E_{c}(t)$ and $E_{e}(t)$ increase with increasing $J$ in Fig.~\ref{figure4}(c) and (d). As a higher-order nonlinear effect, the counter-rotating wave term $\hat{H}_{J,cw}$ between the ions plays an important role in the charging process of the quantum battery.

\begin{figure}[tbph]
\centering\includegraphics[width=9cm]{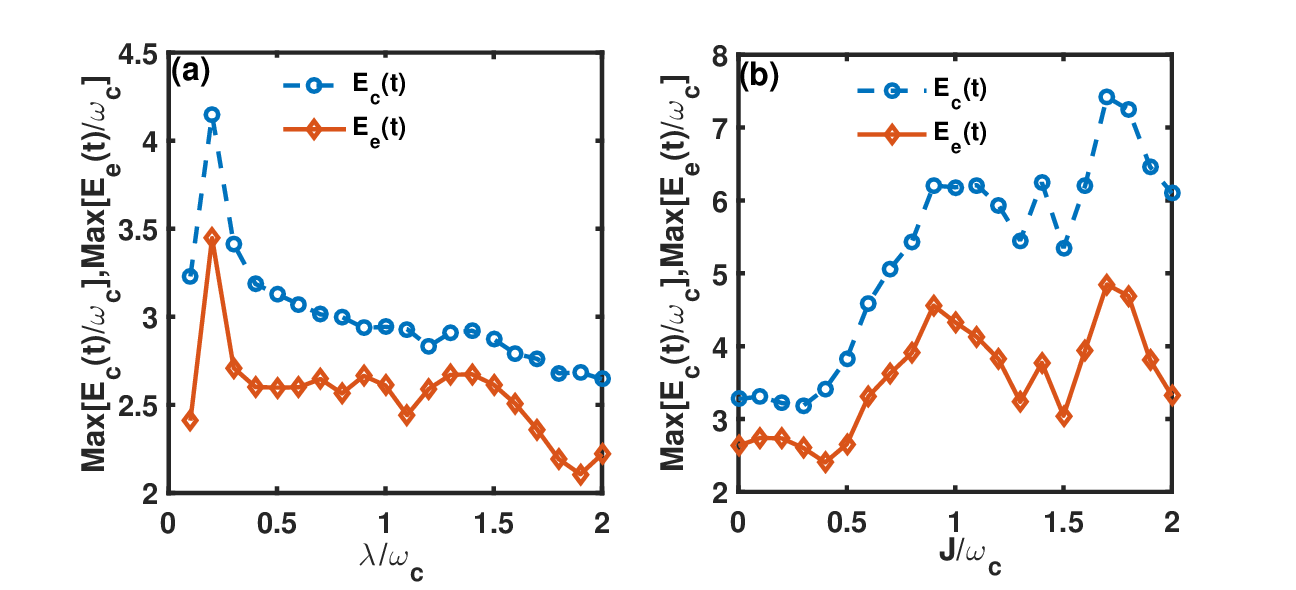}
\caption{(Color online) The change of the maximum of the charging energy $E_{c}(t)$ and the ergotropy $E_{e}(t)$ with the coupling strength $\lambda$ in (a)  for $J=0.3$ and the strength of the hopping interaction $J$ in (b) for $\lambda=0.4$. The time range of the calculation is $\protect\omega _{c}t\in \lbrack 0,30]$ and the exponent $p=3$ is used. }
\label{figure5}
\end{figure}

The maxima of $E_{c}(t)$ and $E_{e}(t)$ at the time range $\protect\omega _{c}t\in \lbrack 0,30]$ are not monotonic
functions of $J$ and $\lambda$, as shown in Fig.~\ref{figure5}. The charging energy is always larger than the ergotropy of the battery. The dependence of the maxima of $E_{c}(t)$ and $E_{e}(t)$ on $\lambda$ is given in Fig.~\ref{figure5}(a). The peak is at $\lambda\approx0.2$, which is the balance point of the competition between the rotating wave term and the counter-rotating wave one in the Hamiltonian $\hat{H}_{ac}$. The counter-rotating wave term $\hat{H}_{ac,cw}$ plays less important role in the charging process for $\lambda<0.2$. That is why in most of the references the rotating wave approximation has been used for studying the quantum dynamics when the value of $\lambda$ is small. Generally, the maxima of $E_{c}(t)$ and $E_{e}(t)$ decrease with increasing $\lambda$ for $\lambda>0.2$. The counter-rotating wave term $\hat{H}_{ac,cw}$ between the ion chain and the mechanical oscillator becomes dominant for larger $\lambda$, which is not conducive for the battery's charging. The dependence of the maxima of $E_{c}(t)$ and $E_{e}(t)$ on $J$ is given in Fig.~\ref{figure5}(b). The change of maxima of $E_{c}(t)$ and $E_{e}(t)$ with $J$ have several extremum points in Fig.~\ref{figure5}(b). Increasing the hopping strength $J$ reduces the low energy levels and expands the whole energy spectrum, but improves the role of the counter-rotating wave term $\hat{H}_{J,cw}$ of the hopping interaction. The former is helpful for the storage of the energy in the ion chain, but the latter breaks the quantum coherence and suppresses the charging. The extremum points come from the competition between both of them in the charging process.

\begin{figure}[tbph]
\centering\includegraphics[width=9cm]{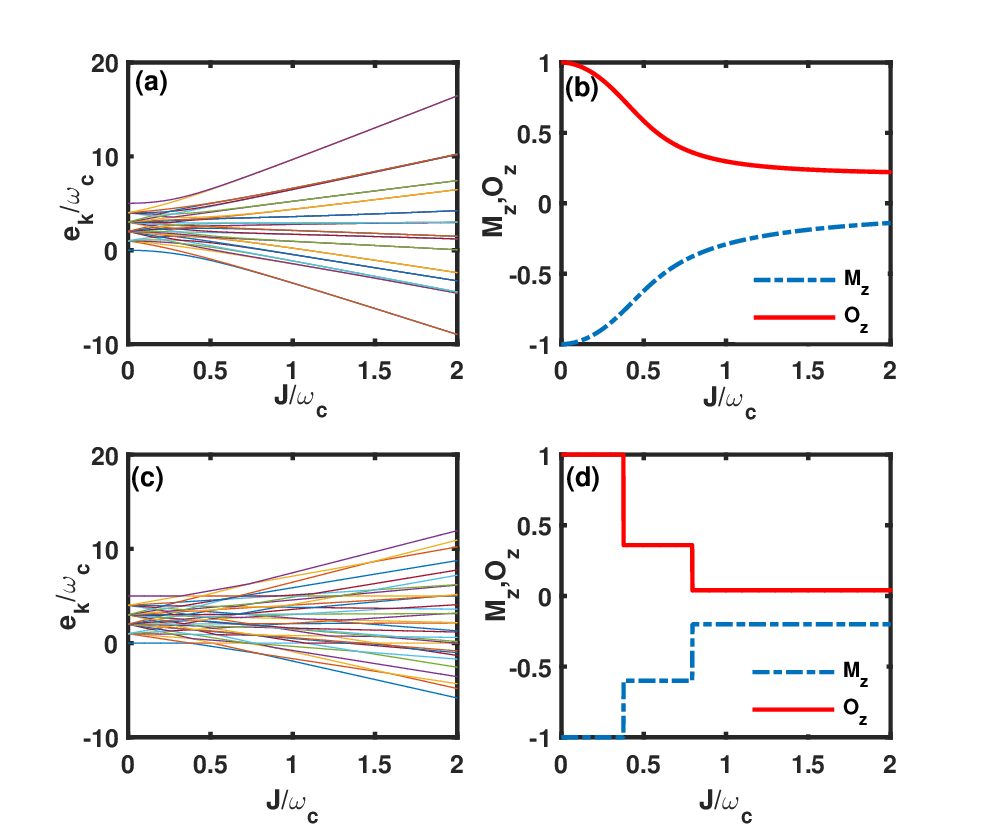}
\caption{(Color online) The change of the eigenvalues $e_{k}$ for $\hat{H}_{a}$ with $J$ in (a, c) and $M_{z}$ and $O_{z}$ in (b, d). The counter-rotating wave term $H_{J,cw}$ is considered in (a, b), but not in (c, d). The exponent $p=3$ is used for both of the cases. }
\label{figure6}
\end{figure}
 A quantum phase transition is supposed to occur in the thermodynamic limit of an infinite number of spins. In Ref.~\cite{AFHS}, the authors have pointed out that there are no quantum phase transition for the system with finite two-level ions by only considering the nearest-neighbor hopping interaction between the ions. Fig.~\ref{figure6}(a) and (b) show the changes of energy spectrum of the ion chain, $O_{z}$ and $M_{z}$ with $J$ for considering the counter-rotating wave term $\hat{H}_{J,cw}$. It is shown that there is no quantum phase transition although the nearest- and non-nearest-neighbor hopping interactions between the five ions are all considered in our model. Fig.~\ref{figure6}(c) and (d) show the change of the energy spectrum, $O_{z}$ and $M_{z}$ with $J$ without considering the counter-rotating wave term $\hat{H}_{J,cw}$. The result about the quantum phase transition is consistent with that given in Ref.~\cite{FQDQ}, which ignores the counter-rotating wave term in the Su-Schrieffer-Heeger quantum battery. That is to say, the counter-rotating wave term between the two-level ions suppresses the quantum coherence and leads to the disappearance of the quantum phase transition.

\begin{figure}[tbph]
\centering\includegraphics[width=9cm]{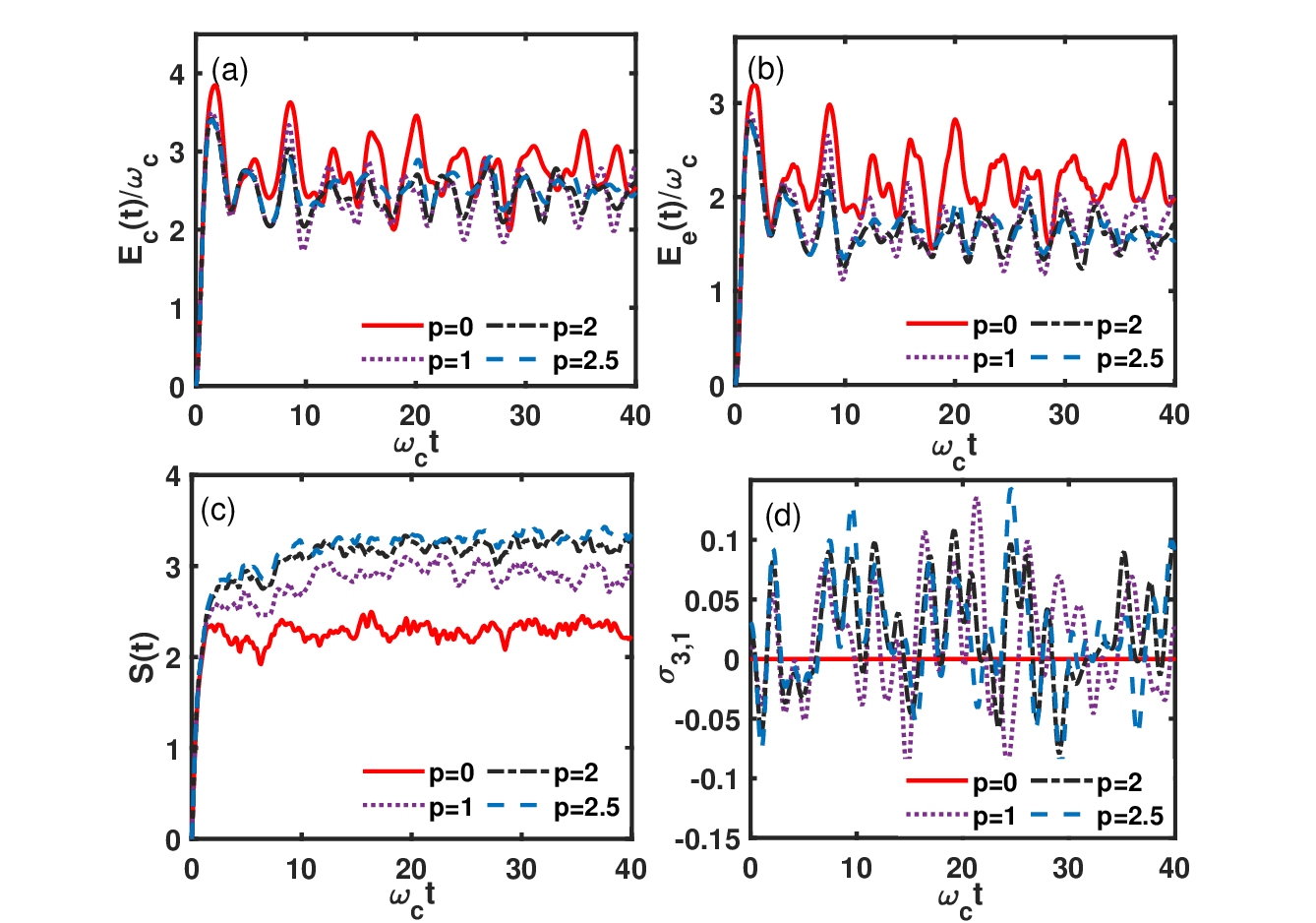}
\caption{(Color online) The charging process of the quantum battery for $p=0$, $1$, $2$, and $2.5$ in the hopping interaction. The evolutions of the charging energy $E_{c}(t)$ in (a), the ergotropy $E_{e}(t)$ in (b), the entropy $S(t)$ in (c), and the difference of the expectation value between the first and third ions $\sigma _{3,1}$ in (d) are given. The values of the other parameters are same with that in Fig.~\ref{figure2}. }
\label{figure7}
\end{figure}
Furthermore,  the influence of the non-nearest-neighbor hopping interaction for
$E_{c}(t)$ and $E_{e}(t)$ is considered. When $J$ is relatively small $(J\ll\omega_{a,c})$,
the effect of the non-nearest-neighbor hopping interaction can be
ignored. However, if $J$ has a relatively large value, the non-nearest-neighbor interaction exerts a discernible influence on the charging process. In particular, the influence also works through the power-law dependence of the distance $\left\vert z_{m}-z_{n}\right\vert$ for the exponent $p$. The hopping interaction has infinite range for $p=0$ and is short-ranged for $p=3$. The power-law dependence in the hopping interaction between the ions is demonstrated in Fig.~\ref{figure7}. By choosing different values of the exponent $p$, the evolutions of the charging energy $E_{c}(t)$, the ergotropy $E_{e}(t)$, the entropy $S(t)$, and the difference of the expectation value between the first and third ions $\sigma _{3,1}$ are given. When $p=0$, the system has the highest symmetry since the hopping interaction doesn't depend on the distance and the time mean of $E_{c}(t)$ and $E_{e}(t)$ is larger than others with $p\neq 0$, as shown in Fig.~\ref{figure7}(a) and (b). The evolution of $E_{c}(t)$ and $E_{e}(t)$ doesn't change very much with the exponent $p$ when $p\geq2$, showing that the change of $p$ has a little effect on the charging ability of the battery. The degree of the entanglement characterized by the final entropy is increased with increasing $p$ until $p\geq2$ from Fig.~\ref{figure7}(c). The dynamics of $\sigma_{3,1}$ in Fig.~\ref{figure7}(d) has similar changing trends for different $p$ except $p=0$. That is to say, the hopping interaction between the ions exhibits the short-ranged behaviours for $p\geq2$ and the result presented in Fig.~\ref{figure2} for $p=3$ serves as a representative example.

Practical quantum batteries must consider the impact of the environment. In the introduction, we present some recent progress on charging of quantum batteries from the view of open systems. For our Dicke-Ising quantum battery of an ion chain, some qualitative arguments about the effects of the environment can be made. As we all know, the environment is either Markovian or non-Markovian. If the resonator is coupled to the Markovian environment and the battery has no interaction with the environment, the final quantum state of the battery is determined by the environment. If the environment is a thermal reservoir, the harmonic oscillator ends up in the thermal state. It is found that if the number of photons in the thermal reservoirs exists, the extractable work of the multiple-spin quantum battery during thermal charging will not be zero and eventually converge to a stable nonzero value~\cite{FZFQ}. Our model also takes into account the coupling between the trapped ions. The nonzero non-diagonal elements in the Hamiltonian suggests the existence of the quantum coherence of the system. The present of thermal phonons in the Markovian environment will make the extractable work of our battery nonzero. Compared to the Markovian environment, some of the information lost to the environment is returned to the system for the non-Markovian environment. The quantum resources (i.e. coherence) in the system are better preserved than in a Markovian environment. In the non-Markovian environment, it is noted that the coefficients of dissipation and fluctuation are positive at some times and negative at others, but their final behaviour tends to zero, the system is similar to one without dissipation and decoherence~\cite{WMZhang2012}. If the role of the non-Markovian environment is taken into account in our model, after a sufficiently long time, the charging dynamics of the system behaves similarly to that of a unitary closed system.

Existing experimental techniques pave the way for the experimental realisation of our charging scheme in two ways. On the one hand, the confinement of the two-level ions can be realised in an ion trap. The method of using a laser to control the strength of the interaction between the ions (including the adjustment of the pairwise power exponent) has been demonstrated in Ref.~\cite{PJBP}. On the other hand, the method to couple the two-level spin system with a mechanical oscillator is presented for example in Ref.~\cite{PBLY}. The coupling strength between the spin and the mechanical mode can be varied by the strength and gradient of the magnetic field. The mechanical oscillator can transfer the vibration energy to the two-level system to realise the charging of the quantum battery. With the effective combination of these two technologies, it is not difficult to implement our proposed charging scheme of the quantum battery.

\section{IV.~Conclusion}
In the present work, we have offered a proposal to realize the quantum batteries in the system of trapped ions, and pointed out that the energy of low-frequency vibration can be converted, stored and utilized. The effects of the hopping interaction between the two-level ions and the coupling interaction between the ions and the external mechanical oscillator on the charging process of the battery are mainly investigated. The complicated roles played by the counter-rotating wave terms in the ultrastrong coupling regime are revealed and discussed by the analytical and numerical calculations. The effect of the environment on the charing dynamics and the available experimental techniques for implementing the charging scheme are discussed in the final remarks. Our work provides a solid foundation for the implementation of quantum batteries and will promote the experimental research in this direction.

\begin{acknowledgments}
\textbf{Acknowledgements}: J.W is supported by Research Foundation of Sichuan Minzu College (Grant No.~KYQD202402C).
P.P and G.Q.L acknowledge the supports from NSF of China (Grant No. 11405100) and the Natural Science Basic Research Plan in Shaanxi Province (Grant Nos.~2019JM-332 and 2020JM-507).
\end{acknowledgments}


\begin{thebibliography}{99}
\bibitem{F.Campaioli2023} F. Campaioli, S. Gherardini, J. Q. Quach, M. Polini, G. M. Andolina,
Colloquium: Quantum batteries,
\href{https://doi.org/10.1103/RevModPhys.96.031001}%
{Rev. Mod. Phys. \textbf{96}, 031001 (2024)}.

\bibitem{RAMF} R. Alicki and M. Fannes, Entanglement boost for extractable
work from ensembles of quantum batteries,
\href{https://doi.org/10.1103/PhysRevE.87.042123}%
{Phys. Rev. E \textbf{87}, 042123 (2013)}.

\bibitem{Allahverdyan2004} A. E. Allahverdyan, R. Balian, and Th. M. Nieuwenhuizen,
Maximal work extraction from finite quantum systems,
\href{https://doi.org/10.1209/epl/i2004-10101-2}%
{Europhys. Lett. \textbf{67}, 565 (2004)}.

\bibitem{KVHM} K. V. Hovhannisyan, M. P. Llobet, M. Huber, and A. Ac\'{\i}n,
Entanglement generation is not necessary for optimal work extraction,
\href{https://doi.org/10.1103/PhysRevLett.111.240401}%
{Phys. Rev. Lett. \textbf{111}, 240401 (2013)}.

\bibitem{FCBS} F. C. Binder, S. Vinjanampathy, K. Modi, and J. Goold,
Quantacell: powerful charging of quantum batteries,
\href{https://doi.org/10.1088/1367-2630/17/7/075015}%
{New J. Phys. \textbf{17}, 075015 (2015)}.

\bibitem{FCFA} F. Campaioli, F. A. Pollock, F. C. Binder, L. Celeri, J.
Goold, S. Vinjanampathy, and K. Modi,
Enhancing the charging power of quantum batteries,
\href{https://doi.org/10.1103/PhysRevLett.118.150601}%
{Phys. Rev. Lett. \textbf{118}, 150601 (2017)}.

\bibitem{TPLJ} T. P. Le, J. Levinsen, K. Modi, M. M. Parish, and F. A.
Pollock, Spin-chain model of a many-body quantum battery,
\href{https://doi.org/10.1103/PhysRevA.97.022106}%
{Phys. Rev. A \textbf{97}, 022106 (2018)}.

\bibitem{GMAM} G. M. Andolina, M. Keck, A. Mari, V. Giovannetti, and M.
Polini, Quantum versus classical many-body batteries,
\href{https://doi.org/10.1103/PhysRevB.99.205437}%
{Phys. Rev. B \textbf{99}, 205437 (2019)}.

\bibitem{W.L.Yang2022} H. L. Shi, S. Ding, Q. K. Wan, X. H. Wang, and W. L. Yang,
Entanglement, coherence, and extractable work in quantum batteries,
\href{https://doi.org/10.1103/PhysRevLett.129.130602}%
{Phys. Rev. Lett.  \textbf{129}, 130602 (2022)}.

\bibitem{F.Mazzoncini2023}
F. Mazzoncini, V. Cavina, G. M. Andolina, P. A. Erdman, and V. Giovannetti,
Optimal control methods for quantum batteries,
\href{https://doi.org/10.1103/PhysRevA.107.032218}%
{Phys. Rev. A \textbf{107}, 032218 (2023)}.


\bibitem{RRRodriguez2024}
R. R. Rodr\'{\i}guez, B. Ahmadi, G Su\'{a}rez, P Mazurek, S Barzanjeh, and P Horodecki, Optimal quantum control of charging quantum batteries,
\href{https://doi.org/10.1088/1367-2630/ad3843}%
{New J. Phys. \textbf{26}, 043004 (2024)}.

\bibitem{KMolmer2019} F. Pirmoradian and K. M{\o}lmer,
Aging of a quantum battery,
\href{https://doi.org/10.1103/PhysRevA.100.043833}%
{Phys. Rev. A \textbf{100}, 043833 (2019)}.

\bibitem{J.Liu2019} J. Liu, D. Segal and G. Hanna,
A loss-free excitonic quantum battery,
\href{https://doi.org/10.1021/acs.jpcc.9b06373}%
{J. Phys. Chem. C \textbf{123}, 18303 (2019)}.

\bibitem{MCarrega2020} M. Carrega, A. Crescente, D. Ferrato and M. Sassetti,
Dissipative dynamics of an open quantum battery
\href{https://doi.org/10.1088/1367-2630/abaa01}%
{New J. Phys. \textbf{22}, 083085 (2020)}.

\bibitem{FHKF} F. H. Kamin, F. T. Tabesh, S. Salimi, F. Kheirandish, and A.
C. Santos,
Non-Markovian effects on charging and self-discharing process of
quantum batteries,
\href{https://doi.org/10.1088/1367-2630/ab9ee2}%
{New J. Phys. \textbf{22}, 003007 (2020)}.

\bibitem{SYBJ} S. Y. Bai and J. H. An,
Floquet engineering to reactivate a dissipative quantum battery,
\href{https://doi.org/10.1103/PhysRevA.102.060201}%
{Phys. Rev. A \textbf{102}, 060201(R) (2020)}.

\bibitem{FZFQ} F. Zhao, F. Q. Dou, and Q. Zhao,
Quantum battery of interacting spins with environmental noise,
\href{https://doi.org/10.1103/PhysRevA.103.033715}%
{Phys. Rev. A \textbf{103}, 033715 (2021)}.

\bibitem{RRRB} R. R. Rodriguez, B. Ahmadi, P. Mazurek, S. Barzanjeh, R.
Alicki, and P. Horodecki, Catalysis in charging quantum batteries,
\href{https://doi.org/10.1103/PhysRevA.107.042419}%
{Phys. Rev. A \textbf{107}, 042419 (2023)}.

\bibitem{D.Rosa2022-1}
J.-Y. Gyhm, D. $\breve{S}$afr$\acute{a}$nek, and D. Rosa,
Quantum charging advantage cannot be extensive without global operations,
\href{https://doi.org/10.1103/PhysRevLett.128.140501}%
{Phys. Rev. Lett. \textbf{128}, 140501 (2022)}.

\bibitem{D.Rosa2020-2}
D. Rossini, G. M. Andolina, D. Rosa, M. Carrega, and M. Polini,
Quantum advantage in the charging process of Sachdev-Ye-Kitaev batteries,
\href{https://doi.org/10.1103/PhysRevLett.125.236402}%
{Phys. Rev. Lett. \textbf{125}, 236402 (2020)}.

\bibitem{Downing2023} C. A. Downing and M. S. Ukhtary,
A quantum battery with quadratic driving,
\href{https://doi.org/10.1038/s42005-023-01439-y}%
{Commun. Phys. \textbf{6}, 322 (2023)}.

\bibitem{C.Shang2024} Z.-G. Lu, G. Tian, X.-Y. L$\ddot{u}$, and C. Shang,
Topological quantum batteries,
\href{https://arxiv.org/abs/2405.03675v3}%
{arXiv: 2405.03675v3}.

\bibitem{CKHu2022}
C.-K. Hu, J. Qiu, P. J. P. Souza, J. Yuan, Y. Zhou, L. Zhang, J. Chu, X. Pan, L. Hu, and J. Li,
Optimal charging of a superconducting quantum battery,
\href{https://doi.org/10.1088/2058-9565/ac8444}%
{Quantum Sci. Technol. \textbf{7}, 045018 (2022)}.

\bibitem{Wenniger2022}
I. Maillette de Buy Wenniger, S. E. Thomas, M. Maffei, S. C. Wein, M. Pont, N. Belabas, S. Prasad, A. Harouri, A. Lema\^{\i}tre, I. Sagnes, N. Somaschi, A. Auff\`{e}ves, P. Senellart,
Experimental analysis of energy transfers between a quantum emitter and light fields,
\href{https://doi.org/10.1103/PhysRevLett.131.260401}%
{Phys. Rev. Lett. \textbf{131}, 260401 (2023)}.

\bibitem{JQQuach2022}
J. Q. Quach, K. E. McGhee, L. Ganzer, D. M. Rouse, B. W. Lovett,
E. M. Gauger, J. Keeling, G. Cerullo, D. G. Lidzey, and T. Virgili,
Superabsorption in an organic microcavity: Toward a quantum battery,
\href{https://doi.org/10.1126/sciadv.abk3160}%
{Science Advances \textbf{8}, 3160 (2022)}.

\bibitem{JJoshi2022}
J. Joshi and T. S. Mahesh,
Experimental investigation of a quantum battery using star-topology NMR spin systems,
\href{https://doi.org/10.1103/PhysRevA.106.042601}%
{Phys. Rev. A \textbf{106}, 042601 (2022)}.

\bibitem{XLDD} X. L. Deng, D. Porras, and J. I. Cirac,
Effective spin quantum phases in systems of trapped ions,
\href{https://doi.org/10.1103/PhysRevA.72.063407}%
{Phys. Rev. A \textbf{72}, 063407 (2005)}.

\bibitem{DPJI} X. L. Deng, D. Porras, and J. I. Cirac,
Quantum phases of interacting phonons in ion traps,
\href{https://doi.org/10.1103/PhysRevA.77.033403}%
{Phys. Rev. A \textbf{77}, 033403 (2008)}.

\bibitem{RBCF} R. Blatt and C. F. Roos,
Quantum simulations with trapped ions,
\href{https://doi.org/10.1038/nphys2252}%
{Nat. Phys. \textbf{8}, 277 (2012)}.

\bibitem{Duan2024}
S. A. Guo, Y. K. Wu, J. Ye, L. Zhang, W. Q. Lian, R. Yao, Y. Wang, R. Y. Yan, Y. J. Yi, Y. L. Xu, B. W. Li, Y. H. Hou, Y. Z. Xu, W. X. Guo, C. Zhang, B. X. Qi, Z. C. Zhou, L. He, and L. M. Duan,
A site-resolved two-dimensional quantum simulator with hundreds of trapped ions,
\href{https://doi.org/10.1038/s41586-024-07459-0}%
{Nature \textbf{630}, 613 (2024)}.

\bibitem{KKim2019}
K. Zhang, J. Thompson, X. Zhang, Y. Shen, Y. Lu, S. Zhang, J. Ma, V. Vedral, M. Gu, and K. Kim,
Modular quantum computation in a trapped ion system,
\href{https://doi.org/10.1038/s41467-019-12643-2}%
{Nat. Commun. \textbf{10}, 4692 (2019)}.

\bibitem{C.Noel2022}
C. Noel, P. Niroula, D. Zhu, A. Risinger, L. Egan, D. Biswas, M. Cetina, A. V. Gorshkov, M. J. Gullans, D. A. Huse, and C. Monroe,
Measurement-induced quantum phases realized in a trapped-ion quantum computer,
\href{https://doi.org/10.1038/s41567-022-01619-7}%
{Nat. Phys. \textbf{18}, 760 (2022)}.

\bibitem{PBLY} P. B. Li, Y. Zhou, W. B. Gao, and F. Nori, Enhancing
spin-phonon and spin-spin interactions using linear resources in a hybrid
quantum system,
\href{https://doi.org/10.1103/PhysRevLett.125.153602}%
{Phys. Rev. Lett. \textbf{125}, 153602 (2020)}.

\bibitem{PJBP} P. Jurcevic, B. P. Lanyon, P. Hauke, C. Hempel, P. Zoller, R.
Blatt, and F. Roos, Quasiparticle engineering and entanglement propagation
in a quantum many-body system,
\href{https://doi.org/10.1038/nature13461}%
{Nature (London) \textbf{511}, 202 (2014)}.

\bibitem{DBRA} D. Braak, Integrability of the Rabi model,
\href{https://doi.org/10.1103/PhysRevLett.107.100401}%
{Phys. Rev. Lett. \textbf{107}, 100401 (2011)}.

\bibitem{VBergholm2019}
V. Bergholm, W. Wieczorek, T. Schulte-Herbr$\ddot{u}$ggen, and M. Keyl,
Optimal control of hybrid optomechanical systems for generating non-classical states of mechanical motion,
\href{https://doi.org/10.1088/2058-9565/ab1682}%
{Quantum Sci. Technol. \textbf{4}, 034001 (2019)}.

\bibitem{PDNation2013}
P. D. Nation,
Nonclassical mechanical states in an optomechanical micromaser analog,
\href{https://doi.org/10.1103/PhysRevA.88.053828}%
{Phys. Rev. A \textbf{88}, 053828 (2013)}.

\bibitem{QZYY} Q. Zheng, Y. Yao, and Y. Li,
Optimal quantum parameter estimation in a pulsed quantum optomechanical system,
\href{https://doi.org/10.1103/PhysRevA.93.013848}%
{Phys. Rev. A \textbf{93}, 013848 (2016)}.

\bibitem{MKounalakis2019}
M. Kounalakis, Y. M. Blanter, and G. A. Steele,
Synthesizing multi-phonon quantum superposition states using flux-mediated three-body interactions with superconducting qubits, \href{https://doi.org/10.1038/s41534-019-0219-y}%
{npj Quantum Information \textbf{5}, 100 (2019)}.

\bibitem{H.Tan2014}
H. Tan,
Deterministic quantum superpositions and Fock states of mechanical oscillators via quantum interference in single-photon cavity optomechanics,
\href{https://doi.org/10.1103/PhysRevA.89.053829}%
{Phys. Rev. A \textbf{89}, 053829 (2014)}.

\bibitem{H.Xie2019}
H. Xie, X. Shang, C.-G. Liao, Z.-H. Chen, and X.-M. Lin,
Macroscopic superposition states of a mechanical oscillator in an optomechanical system with quadratic coupling,
\href{https://doi.org/10.1103/PhysRevA.100.033803}%
{Phys. Rev. A \textbf{100}, 033803 (2019)}.

\bibitem{JJing2024}
C.-Y Zhang and J. Jing,
Generating Fock-state superpositions from coherent states by selective measurement, \href{https://doi.org/10.1103/PhysRevA.110.042421}%
{Phys. Rev. A \textbf{110}, 042421 (2024)}.

\bibitem{Y.Chu2018}
Y. Chu, P. Kharel, T. Yoon, L. Frunzio, P. T. Rakich, and R. J. Schoelkopf,
Creation and control of multi-phonon Fock states in a bulk acoustic-wave resonator, \href{https://doi.org/10.1038/s41586-018-0717-7}%
{Nature (London) \textbf{563}, 666 (2018)}.

\bibitem{DFVJ} D. F. V. James,
Quantum dynamics of cold trapped ions with application to quantum computation,
\href{https://doi.org/10.1007/s003400050373}%
{Appl. Phys. B. \textbf{66}, 181 (1998)}.

\bibitem{DFMC} D. Ferraro, M. Campisi, G. M. Andolina, V. Pellegrini, and M.
Polini, High-power collective charging of a solid-state quantum battery,
\href{https://doi.org/10.1103/PhysRevLett.120.117702}%
{Phys. Rev. Lett. \textbf{120}, 117702 (2018)}.

\bibitem{CHBD} C. H. Bennett, D. P. Divincenzo, J. A. Smolin, and W. K.
Wootters, Mixed-state entanglement and quantum error correction,
\href{https://doi.org/10.1103/PhysRevA.54.3824}%
{Phys. Rev. A \textbf{54}, 3824 (1996)}.

\bibitem{FQDQ} F. Zhao, F. Q. Dou, and Q. Zhao,
Charging performance of the Su-Schrieffer-Heeger quantum battery,
\href{https://doi.org/10.1103/PhysRevResearch.4.013172}%
{Phys. Rev. Research \textbf{4}, 013172 (2022)}.

\bibitem{XTZH} Q. T. Xie, H. H. Zhong, M. T. Batchelor, and C. H. Lee,
The quantum Rabi model: solution and dynamics,
\href{https://doi.org/10.1088/1751-8121/aa5a65}%
{J. Phys. A: Math. Theor, \textbf{50}, 113001 (2017)}.

\bibitem{JCGR} J. Casanova, G. Romero, I. Lizuain, J. J. G. Ripoll, and E.
Solano, Deep strong coupling regime of the Jaynes-Cummings model,
\href{https://doi.org/10.1103/PhysRevLett.105.263603}%
{Phys. Rev. Lett. \textbf{105}, 263603 (2010)}.

\bibitem{JFHC} J. F. Huang and C. K. Law, Phase-kicked control of
counter-rotating interactions in the quantum Rabi model,
\href{https://doi.org/10.1103/PhysRevA.91.023806}%
{Phys. Rev. A \textbf{91}, 023806 (2015)}.

\bibitem{HZSY} H. Zheng, S. Y. Zhu, and M. S. Zubariy, Quantum Zeno and
anti-Zeno effects: without the rotating-wave approximation,
\href{https://doi.org/10.1103/PhysRevLett.101.200404}%
{Phys. Rev. Lett. \textbf{101}, 200404 (2008)}.

\bibitem{U.R.Fischer2024}
J.-Y. Gyhm and U. R. Fischer,
Beneficial and detrimental entanglement for quantum battery charging,
\href{https://doi.org/10.1116/5.0184903}%
{AVS Quantum Sci. \textbf{6}, 012001 (2024)}.

\bibitem{DLYang2023} D. L. Yang, F. M. Yang, and F. Q. Dou,
Three-level Dicke quantum battery,
\href{https://doi.org/10.1103/PhysRevB.109.235432}%
{Phys. Rev. B \textbf{109}, 235432 (2024)}.

\bibitem{AFHS} A. Friedenauer, H. Schmitz, J. T. Glueckert, D. Porras, and
T. Schaetz, Simulating a quantum magnet with trapped ions,
\href{https://doi.org/10.1038/nphys1032}%
{Nat. Phys. \textbf{4}, 757 (2008)}.

\bibitem{WMZhang2012} W. M. Zhang, P. Y. Lo, H. N. Xiong, M. W. Y. Tu, and F. Nori,
General non-Markovian dynamics of open quantum systems,
\href{https://doi.org/10.1103/PhysRevLett.109.170402}%
{Phys. Rev. Lett. \textbf{109}, 170402 (2012)}.


\end{thebibliography}
\end{document}